\documentclass{article}
\usepackage[T1]{fontenc} 
\usepackage[utf8]{inputenc} 
\usepackage{ismir}

\usepackage{amsmath}
\usepackage{amsfonts}
\usepackage{amssymb}
\usepackage{amsthm}
\usepackage{mathabx}
\usepackage{mathtools}
\usepackage{stmaryrd}

\usepackage{upgreek}

\usepackage{paralist}

\usepackage{cite}
\usepackage{url}

\usepackage{graphicx}

\usepackage{color}
\usepackage[table]{xcolor}

\usepackage{siunitx}
\usepackage{tabularx}
\usepackage{booktabs}

\newif\ifdoubleblind

\usepackage{lineno}

\usepackage{marginnote}

\newcommand{\ackedalex}[1]{}

\usepackage{collcell}
\usepackage{colortbl}
\usepackage{tikz}
\usepackage{lipsum}

\usepackage{xspace}

\newcommand{\etal}{et al.\@\xspace}
\newcommand{\etalc}{et~al.\@~}

\title{A Stem-Agnostic Single-Decoder System for\\ Music Source Separation Beyond Four Stems}

\oneauthor
{Karn N.\ Watcharasupat \qquad Alexander Lerch}
{Music Informatics Group, Georgia Institute of Technology, Atlanta, GA, USA\\
{\tt \{kwatcharasupat, alexander.lerch\}@gatech.edu}}

\def\authorname{K.\ N.\ Watcharasupat, and A.\ Lerch}

\usepackage[bookmarks=false,%
    pdfauthor={\authorname},%
    pdfsubject={\papersubject},%
    hidelinks]{hyperref}

\usepackage{cleveref}
\crefname{figure}{fig.\!}{figs.\!}
\Crefname{figure}{Fig.\!}{Figs.\!}

\begin{document}

\maketitle

\begin{abstract}
Despite significant recent progress across multiple subtasks of audio source separation, few music source separation systems support separation beyond the four-stem vocals, drums, bass, and other (\textsc{vdbo}) setup. Of the very few current systems that support source separation beyond this setup, most continue to rely on an inflexible decoder setup that can only support a fixed pre-defined set of stems. Increasing stem support in these inflexible systems correspondingly requires increasing computational complexity, rendering extensions of these systems computationally infeasible for long-tail instruments. We propose Banquet,
a system that allows source separation of multiple stems using just one decoder. A bandsplit source separation model is extended to work in a query-based setup in tandem with a music instrument recognition PaSST model. On the MoisesDB dataset, Banquet~---~at only 24.9~M trainable parameters~---~performed on par with or better than the significantly more complex 6-stem Hybrid Transformer Demucs.
The query-based setup allows for the separation of narrow instrument classes such as clean acoustic guitars, and can be successfully applied to the extraction of less common stems such as reeds and organs.
\end{abstract}
\section{Introduction}\label{sec:introduction}

Music Source Separation (MSS) is the task of separating a musical audio mixture into its constituent components, commonly referred to as stems. 
The releases~of DSD100~\cite{Liutkus20172016SignalSeparation} and MUSDB18  \cite{Rafii2017MUSDB18CorpusMusic, Rafii2019MUSDB18HQUncompressedVersion}, both being four-stem MSS datasets, have defined a de-facto standard, with nearly every major work since relying on the four-stem \textit{vocals}, \textit{bass}, \textit{drum}, and \textit{others} (\textsc{vdbo}) setup
\cite{Uhlich2017ImprovingMusicSource, Takahashi2017MultiScaleMultibandDensenets, Stoter20182018SignalSeparation, Stoller2018WaveUNetMultiScaleNeural, Stoter2019OpenUnmixReferenceImplementation, Defossez2019DemucsDeepExtractor, Defossez2019MusicSourceSeparation, Hennequin2020SpleeterFastEfficient, Takahashi2021D3NetDenselyConnected, Kong2021DecouplingMagnitudePhase, Defossez2021HybridSpectrogramWaveform, Kim2021KUIELabMDXNetTwoStreamNeural, Mitsufuji2022MusicDemixingChallenge, Luo2023MusicSourceSeparation, Rouard2023HybridTransformersMusic, Lu2023MusicSourceSeparation}. While this has significantly improved the comparability and reproducibility of the task, it has also disproportionately favored the \textsc{vdbo} setup. Very few works have tackled MSS beyond the \textsc{vdbo} setup, each relying on datasets with significant limitations: Wang et al.~\cite{Wang2022FewShotMusicalSource} relied on MedleyDB \cite{Bittner2014MedleyDBMultitrackDataset, Bittner2016MedleyDBNewData}, whose stem ontology is somewhat unfriendly to source separation, Manilow et al.~\cite{Manilow2020HierarchicalMusicalInstrument} relied on the synthetically generated Slakh dataset \cite{Manilow2019CuttingMusicSource}, and others relied on proprietary data inaccessible to other research groups \cite{Hennequin2020SpleeterFastEfficient, Rouard2023HybridTransformersMusic}, limiting reproducibility.
The recently released MoisesDB \cite{Pereira2023MoisesDBDatasetSource}, a multitrack source separation dataset, attempts to address these limitations, 
particularly in terms of stem availability and taxonomy. This aims at broadening the task 
beyond \textsc{vdbo} based on publicly available data. 
However, to the best of our knowledge, while MoisesDB was used in the 2023 Sound Demixing Challenge (SDX) \cite{Fabbro2024SoundDemixingChallenge}, no published system has utilized MoisesDB for source separation beyond \textsc{vdbo} yet.

In this work, we propose Banquet,\footnote{Banquet is a portmanteau of \textbf{Que}ry-based \textbf{Band}i\textbf{t}. Code available at \href{https://github.com/kwatcharasupat/query-bandit}{github.com/kwatcharasupat/query-bandit}. Last accessed 24 July 2024.} a query-based source separation model that can separate an arbitrary number of stems using just one set of stem-agnostic encoder and decoder, and a pre-trained feature extractor \cite{Koutini2022EfficientTrainingAudio}. Our model was adapted from the 
cinematic audio source separation Bandit model \cite{Watcharasupat2023GeneralizedBandsplitNeural}, which was in turn adapted from the music source separation Bandsplit RNN model~\cite{Luo2023MusicSourceSeparation}. Bandit significantly reduces the complexity of Bandsplit RNN by adopting a common-encoder approach with stem-specific decoders. In this work, we take the complexity reduction further by switching to a query-based setup, using only one decoder shared amongst all possible stems. Performance evaluation on MoisesDB demonstrated separation performance above oracle for drum and bass, state-of-the-art for guitar and piano, and at least \SI{7.4}{dB} SNR for vocals. Our system additionally provided support for fine-level stem extraction currently available only in a few MSS systems.



\section{Related Work}



Nearly every major MSS works since 2017 have relied on the \textsc{vdbo} setup. Early systems \cite{Uhlich2017ImprovingMusicSource, Takahashi2018MMDenseLSTMEfficientCombination, Stoter20182018SignalSeparation}, including Open-Unmix \cite{Stoter2019OpenUnmixReferenceImplementation}, were usually Time-Frequency (TF) masking models with LSTM forming the core of the systems, with some experimenting with densely-connected convolutional systems \cite{Takahashi2017MultiScaleMultibandDensenets, Takahashi2021D3NetDenselyConnected}. Beginning with Wave-U-Net \cite{Stoller2018WaveUNetMultiScaleNeural}, the U-Net architecture became a popular choice for MSS, with notable models such as Demucs \cite{Defossez2019DemucsDeepExtractor, Defossez2019MusicSourceSeparation, Defossez2021HybridSpectrogramWaveform, Rouard2023HybridTransformersMusic}, Spleeter~\cite{Hennequin2020SpleeterFastEfficient}, ByteSep \cite{Kong2021DecouplingMagnitudePhase}, and KUIELab-MDX-Net \cite{Kim2021KUIELabMDXNetTwoStreamNeural} all being some variations of a U-Net. More recently, Bandsplit RNN \cite{Luo2023MusicSourceSeparation} became one of the few state-of-the-art systems to not rely on a U-Net setup. This was followed by the Bandsplit RoPE Transformer model \cite{Lu2023MusicSourceSeparation} topping the leaderboard of SDX 2023 \cite{Fabbro2024SoundDemixingChallenge}. Of existing open-source systems, very few offer separation functionality beyond the \textsc{vdbo} setup. Spleeter \cite{Hennequin2020SpleeterFastEfficient} supports 5-stem separation with \textsc{vdbo} and piano. HT-Demucs \cite{Rouard2023HybridTransformersMusic} supports a 6-stem setup with \textsc{vdbo}, piano, and guitar. 

\subsection{Conditional source separation}

The systems mentioned above were mostly designed 
with either stem-specific models, stem-specific decoders, or a shared decoder with predetermined outputs. As a result, these systems are not particularly amenable to the addition of new stems, especially if these new stems have limited data availability. Below we review some of the common approaches for conditional source separation that may be useful for extending existing systems beyond \textsc{vdbo}. 

Meseguer-Brocal and Peeters \cite{Meseguer-Brocal2019ConditionedUNetIntroducingControl} were likely amongst the first to attempt a conditioned U-Net for source separation using a single decoder. 
They used multiple feature-wise linear modulation (FiLM) \cite{Perez2017FiLMVisualReasoning} layers within the encoder to perform MSS in a \textsc{vdbo} setup. Slizovskaia \etalc{}\cite{Slizovskaia2021ConditionedSourceSeparation} used a similar setup with FiLMs either throughout the encoder, at the bottleneck layer, or at the final decoder layer. The systems in \cite{Slizovskaia2021ConditionedSourceSeparation} were tested on the 13-instrument URMP dataset \cite{Li2019CreatingMultitrackClassical}, with up to 4 active instruments in any recording, but all performed poorly in terms of mean signal-to-distortion ratio (SDR). 
Lin \etalc{}\cite{Lin2021UnifiedModelZeroshot} proposed a joint separation-transcription U-Net system, which performed well for string and brass instruments in URMP, but struggled on woodwind instruments. The system in \cite{Lin2021UnifiedModelZeroshot} used FiLMs throughout the encoder with a query embedding from another convolutional model, and across all skip connections with transcription embeddings. 

Lee \etalc{}\cite{Lee2019AudioQuerybasedMusic} proposed a U-Net with two methods of less aggressive conditioning with examples beyond \textsc{vdbo}, but only provided objective results for a \textsc{vdbo} setup on MUSDB18. Wang \etalc{}\cite{Wang2022FewShotMusicalSource} also proposed a U-Net, with FiLM conditioning only at the bottleneck layer. The system in \cite{Wang2022FewShotMusicalSource} was able to support a substantial number of stems beyond \textsc{vdbo} with the caveat that its reported performance is significantly below contemporary models for \textsc{vdbo} stems. Gfeller \etalc{}\cite{Gfeller2021OneShotConditionalAudio} utilized a FiLM-conditioned wave-to-wave U-Net to perform one-shot conditional audio filtering. Similar approaches were also adopted in Choi \etalc{}\cite{Choi2021LaSAFTLatentSource} and Jeong \etalc{}\cite{Jeong2021LightSAFTLightweightLatent} for MSS, in Chen \etalc{}\cite{Chen2022ZeroshotAudioSource} for source activity-queried separation, in Kong \etalc{}\cite{Kong2023UniversalSourceSeparation} for universal source separation, and in Liu \etalc{}\cite{Liu2022SeparateWhatYou, Liu2023SeparateAnythingYou} for language-queried source separation. These works \cite{Gfeller2021OneShotConditionalAudio,Choi2021LaSAFTLatentSource,Jeong2021LightSAFTLightweightLatent,Chen2022ZeroshotAudioSource,Kong2023UniversalSourceSeparation,Liu2022SeparateWhatYou, Liu2023SeparateAnythingYou} applied FiLM or generalizations thereof to nearly every single layer of the network, significantly increasing the computational complexity of the system. We surmise that the apparent need for multiple conditioning in a U-Net is probably due to the nature of its information flow \cite{Lee2021InformationFlowUNets}, which may require a significant number of information streams to be conditioned to achieve acceptable performance. 

In a different direction, source separation systems relying on audio embedding ``distances'' have also been developed, notably with Le Roux \etal{} in \cite{Seetharaman2019ClassconditionalEmbeddingsMusic, Manilow2020HierarchicalMusicalInstrument, Petermann2023HyperbolicAudioSource}. 
In 2018, Kumar \etalc{}\cite{Kumar2018MusicSourceActivity} presented an early work using Euclidean audio embedding distance from a ``query'' embedding to inform music source separation. A similar system using a Gaussian mixture model posterior in lieu of standard distance was proposed in \cite{Seetharaman2019ClassconditionalEmbeddingsMusic}. 
Hierarchical masking \cite{Manilow2020HierarchicalMusicalInstrument} was later utilized to allow the extraction of stems at multiple levels of specificity. More recently, source separation systems with audio embedding in a low-dimensional hyperbolic space have been developed to allow music \cite{Petermann2023HyperbolicAudioSource} and speech \cite{Petermann2024HyperbolicDistanceBasedSpeech} source separation with some degrees of control on the specificity of the extraction. Uniquely, Samuel \etalc{}\cite{Samuel2020MetaLearningExtractorsMusic} proposed a network-generating network approach for instrument-conditioned source separation.

\section{Proposed system}

\begin{figure}
    \centering
    \includegraphics[alt={A flowchart showing the architectural overview of the proposed Banquet system, including the query signal encoding, mixture signal encoding, query adaptor, and estimated stem decoding.}, width=0.8\columnwidth]{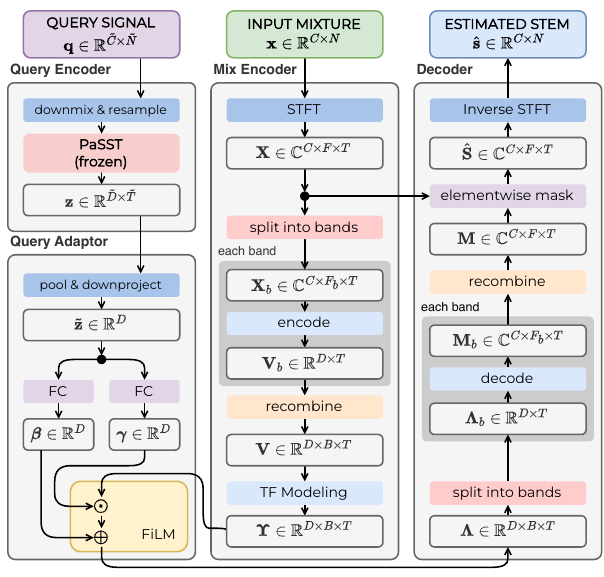}
    \vskip-2.5ex
    \caption{Overview of the Banquet System. }
    \label{fig:system}
\end{figure}

The overview of the proposed Banquet system is shown in \Cref{fig:system}. The system is a single-encoder single-decoder adaptation of Bandit \cite{Watcharasupat2023GeneralizedBandsplitNeural}, that takes in a mixture signal $\mathbf{x}$ and a query signal $\mathbf{q}$, and extracts a stem estimate $\hat{\mathbf{s}}$ from the mixture signal of the ``same'' stem type as the query signal using a complex-valued TF mask. This is done by 
\begin{inparaenum}[(i)]
    \item encoding the mixture into a subband-level time-varying embedding tensor $\mathbf{\Upsilon}$, 
    \item encoding the query into a single-vector representation $\tilde{\mathbf{z}}$,
    \item adapting the mixture embedding, conditioned on the query, into a stem-specific embedding $\mathbf{\Lambda} = \mathcal{Q}(\mathbf{\Upsilon}; \tilde{\mathbf{z}})$, then 
    \item decoding the $\mathbf{\Lambda}$ to a TF mask $\mathbf{M}$ that is used to obtain the source estimate. 
\end{inparaenum}

\protect\ackedalex{is it intentional that the input informs the query and not the other way around? While that might be equivalent, it seems a bit weird}

\subsection{Bandit encoder} 
\ackedalex{This seems like an awfully long subsection for something that has already been published elsewhere. Isn't only the parametrization new here?}
The encoder module of the system used in this work is the \textit{musical} variant of the Bandit encoder, with $B = 64$ bands. Specifically, given an input mixture $\mathbf{x} \in \mathbb{R}^{C \times N}$ with $C$ channels and $N$ samples, a short-time Fourier transform (STFT) of $\mathbf{x}$ is computed to obtain $\mathbf{X} \in \mathbb{C}^{C \times F \times T}$ with a frame size of $N_{\text{FFT}} = 2(F-1) = 2048$ and \SI{75}{\percent} overlap.  The STFT is then split into overlapping subbands as detailed in \cite{Watcharasupat2023GeneralizedBandsplitNeural}.
Each of the subbands is then viewed as a real-valued tensor in $\mathbb{R}^{2CF \times T}$, passed through a layer norm and an affine transformation with $D=128$ neurons to obtain $\mathbf{V}_b \in \mathbb{R}^{D \times T}$. These tensors are then stacked to obtain $\mathbf{V} \in \mathbb{R}^{D \times B \times T}$. TF modeling is then applied on $\mathbf{V}$ to obtain $\mathbf{\Upsilon}$ using 8~pairs of residual gated recurrent units (GRUs), the first of each pair operating along the time axis and the second along the band axis. 

Note that this TF modeling is the only part of the model that is recurrent across either the time or the subband axes. The rest of the encoder and the decoder operate in a subband-wise manner identically for any time frame.

\subsection{Query encoding}

To obtain the query embedding, a PaSST model \cite{Koutini2022EfficientTrainingAudio} trained on the OpenMIC-2018 dataset \cite{Humphrey2018OpenMIC2018OpenDataset} is used. The 20 instruments in OpenMIC span all coarse-level classes of MoisesDB, except \textit{other}. For compatibility, each query signal is downmixed to mono and downsampled to \SI{32}{\kilo\hertz} before being fed to PaSST. Although the query feature extractor could, in theory, be jointly trained with the rest of the system, preliminary experiments showed that this can result in considerable instability during training, especially if the query feature extractor is not at least pretrained. Due to the size and complexity of PaSST, the query feature extractor is fully frozen in this work. The embedding from the PaSST variant used is a time series with a feature dimension of $\tilde{D}=784$. The embedding is averaged over time and linearly down-projected to obtain $\tilde{\mathbf{z}} \in \mathbb{R}^D$.

\subsection{Query-based adaptation}

In the original Bandit system \cite{Watcharasupat2023GeneralizedBandsplitNeural}, each stem was estimated through a dedicated decoder. As a result, $\mathbf{\Upsilon}$ typically contains information from all stems, with most of the ``separation'' occurring within each of the decoders. This is evident in the fact that the encoder of a Bandit system trained on the cinematic audio Divide and Remaster (DnR) dataset \cite{Petermann2022CocktailForkProblem} could be successfully used in a 4-stem MSS on the MUSDB18-HQ dataset \cite{Rafii2017MUSDB18CorpusMusic} with separation quality on par with Open-Unmix \cite{Watcharasupat2023GeneralizedBandsplitNeural}.

In this work, only a single decoder is responsible for mask estimation for any stem. As a result, the query-based adaptation $\mathcal{Q} \colon (\mathbb{R}^{D \times B \times T}, \mathbb{R}^{D}) \mapsto \mathbb{R}^{D \times B \times T}$ has an important role in filtering out irrelevant information from $\mathbf{\Upsilon}$, or at least ``hinting'' to the decoder the nature of the target stem. A single FiLM layer is used to map from the mixture embedding to the stem-specific embedding, that is,
\begin{equation}
    \mathbf{\Lambda}[d, b, t] = \bm{\upgamma}[d] \cdot \mathbf{\Upsilon}[d, b, t] + \bm{\upbeta}[d],\ \forall d,b,t,
\end{equation}
where modulating variables $\bm{\upgamma}, \bm{\beta} \in \mathbb{R}^{D}$ are obtained from a two-layer nonlinear affine map of $\mathbf{\tilde{z}}$. This is similar to the conditioning method used in \cite{Wang2022FewShotMusicalSource}.

Crucially, note that the modulating variables are not subband-specific. Due to the nature of the TF modeling module within the encoder, features of $\mathbf{\Upsilon}$ are already aligned across subbands and time frames. Moreover, BSRNN-like models only contain one stream of information flow, with a clear bottleneck, thus lending itself to the global conditioning mechanism significantly more than, for example, U-Net-style models in \cite{Wang2022FewShotMusicalSource, Liu2022SeparateWhatYou, Liu2023SeparateAnythingYou}. 

The use of embedding-based query, as opposed to one-hot class-based query, provides significant practical flexibility in adding new instruments as data become available or in adjusting the level of specificity in the querying, as these can be done via finetuning with no architectural changes to the model. Moreover, class-based query can be emulated in an embedding-based system but not vice versa.

\subsection{Bandit decoder}

The decoder used is identical in structure to that in \cite{Watcharasupat2023GeneralizedBandsplitNeural}.
The major difference is that there is only one stem-agnostic decoder. Given a conditioned embedding tensor $\mathbf{\Lambda}$, the embedding tensor is split into subband-level representation $\mathbf{\Lambda}_b = \mathbf{\Lambda}[\colon, b, \colon]$. Each $\mathbf{\Lambda}_b$ is passed through a layer norm and a gated linear unit (GLU) to obtain a real-valued tensor $\mathbb{R}^{2CF_b \times T}$ which is then viewed as a complex-valued tensor $\mathbf{M}_b \in \mathbb{C}^{C \times F_b \times T}$. Frequency-domain overlap-add is then applied to obtain the full-band mask using
\begin{align}
    \mathbf{M}[c, f, t] = \sum_{b=0}^{B-1} \frac{\mathbf{W}[b, f] \cdot \mathbf{M}_b[ c, f- \min\mathfrak{F}_b, t]}{\sum_{k=0}^{B-1} \mathbf{W}[k, f]}
\end{align}
Finally, the source estimates are then obtained using elementwise masking $\hat{\mathbf{S}} = \mathbf{X} \circ \mathbf{M}$.

\subsection{Loss function}
The loss function used in this work is the multichannel version of the L1SNR loss proposed in \cite{Watcharasupat2023GeneralizedBandsplitNeural}.
The contribution for each sample of the loss function is given by
\begin{align}
    \mathcal{L}(\hat{\mathbf{s}}; \mathbf{s}) &= \mathcal{D}(\hat{\mathbf{s}}; \mathbf{s}) + \mathcal{D}(\Re\hat{\mathbf{S}}; \Re\mathbf{S}) + \mathcal{D}(\Im\hat{\mathbf{S}}; \mathbf{S}),\\
     \mathcal{D}(\hat{\mathbf{y}}; \mathbf{y})
    &= 10 \log_{10} \frac{\|\operatorname{vec}(\hat{\mathbf{y}} - \mathbf{y})\|_1  + \epsilon}{\|\operatorname{vec}(\mathbf{y})\|_1  + \epsilon},
\end{align}
where $\hat{\mathbf{s}} = \operatorname{iSTFT}(\hat{\mathbf{S}})$, $\mathbf{s}$ and $\mathbf{S}$ are defined similarly for the ground truth, $\operatorname{vec}(\cdot)$ is the vectorization operator, and $\epsilon = 10^{-3}$ for stability.
\ackedalex{Why do you need all this detail if it's just the same already published loss function? Does this impact any discussions later?} 

\section{Data and experimental setup}

This work utilizes the MoisesDB dataset \cite{Pereira2023MoisesDBDatasetSource}, which consists of 240 songs from 47 artists, in stereo format at \SI{44.1}{\kilo\hertz}.  MoisesDB defined their stem ontology with more than 30 fine-level classes, which are then grouped into 11 coarse-level classes \cite[Table 2]{Pereira2023MoisesDBDatasetSource}. Due to the lack of official splits for MoisesDB, we performed a five-fold split\footnote{The splits are available in the repository. Note that not all stems contain a sufficient number of data points to be split into a five-fold validation setup. As a result, some stems are only present in a subset of folds. } on the dataset stratified by genres. The first three splits are used as the training set, the fourth as the validation set, and the last as the test set. 

\subsection{Query extraction}

\begin{figure}[t]
    \centering
    \includegraphics[alt={A t-SNE plot of the PaSST embeddings of the query signals. Embeddings form clusters mostly based on instrument labels, but with some overlaps, especially amongst keyboard instruments.}, width=.9\linewidth]{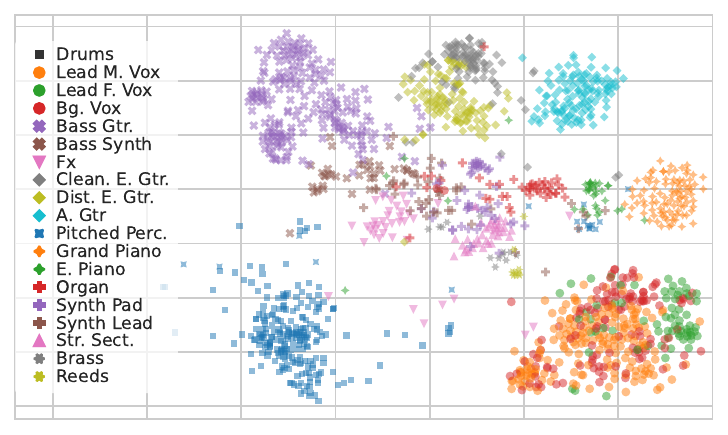}
    \vspace{-2ex}
    \caption{A t-SNE plot of the PaSST embeddings of the query signals. Stems from the same coarse-level grouping, as defined by \cite{Pereira2023MoisesDBDatasetSource}, share the same symbol.}
    \label{fig:tsne}
\end{figure}

For each possible stem of each song, a 10-second chunk of the clean audio of the same stem
is extracted as the query signal. This is done by computing a time series of onset strength for each stem and then aggregating the mean onset strength for each 10-second sliding window with a hop size of 512 samples.
The 10-second window with the strongest average onset is taken as the query signal.
A t-SNE plot of the query embedding is shown in \Cref{fig:tsne}. While clusters can be clearly seen amongst related stems, it can also be seen that there are varying degrees of non-separability of the embedding between fine-level stems.

\subsection{Training}

Each model was trained using an NVIDIA H100 GPU (80 GB) for up to 150 epochs, unless otherwise stated. A training epoch consists of 8192 mixture-query pairs, with a batch size of 4. We used Adam optimizer with an initial learning rate of \num{e-3} and a decay factor of 0.98 per epoch. 

In the default sampling strategy, a random song is chosen, a random trainable stem for that song is chosen as the target stem, then a random chunk of \SI{6}{\second} is chosen. If the current target chunk has an RMS below \SI{-36}{dBFS}, a new random chunk is chosen for up to 10 more trials. Otherwise, the threshold is dropped to \SI{-48}{dBFS} for another 10 trials. If a suitable chunk is still not found, the next random chunk is chosen regardless of RMS. A pre-extracted query of the same stem is then randomly chosen from the available pool of songs, including the song of the mixture.

\subsection{Testing and inference}

During testing and inference, each track is split into 6-s segments with a hop size of \SI{0.5}{\second}, as per \cite{Luo2023MusicSourceSeparation}.
The estimated stems were then reconstructed into a full track using time-domain overlap-add with a Hann window. The Banquet models are tested in two scenarios: one using a query from a different song, and another using a query from the same song (SSQ). In different-song querying, the query song for each stem is randomly chosen from another song within the test split that contains the stem. When possible, the query song is chosen so that it is from the same genre as the mixture song but from a different artist. Otherwise, a song from any genre with a different artist is chosen.

\subsection{Evaluation metric}

In this work, we report the full-track multichannel signal-to-noise ratio (SNR)\footnote{Signal-to-interference ratio (SIR) and signal-to-artifact ratio (SAR) were not computed as the number of the constituent stems can be large, making the required subspace projection intractable and/or unreliable. It is also unclear if coarse-level ground truth or fine-level ground truth should be used for such a projection. See \cite{Vincent2006PerformanceMeasurementBlind, LeRoux2019SDRHalfbakedWell, Scheibler2022SDRMediumRare} for background.} as the main metric. Specifically, for a test signal $\hat{\mathbf{s}}$ and a reference signal $\mathbf{s}$, both in $\mathbb{R}^{C \times N}$, the SNR is computed by
\begin{align}
    \operatorname{SNR}(\hat{\mathbf{y}}; \mathbf{y}) = 10 \log_{10} \left({\| \mathbf{s}\|_F^2}/{\|\hat{\mathbf{s}} - \mathbf{s}\|_F^2}\right).
\end{align} 

\section{Results and discussion}

In this section, we provide the results and discussion of our experiments. \Cref{sec:results/pretrained} discusses pretraining of the Bandit/Banquet encoder. \Cref{sec:results/vdb} trials the use of the query-based setup on a subset of vocals, drums, and bass stems. \Cref{sec:results/vdbgp} extends the system to include fine-level stems from \textit{guitar} and \textit{piano} families. Finally, \Cref{sec:results/ev} attempts to perform extraction on all possible fine-level stems with sufficient data.

\subsection{Encoder pretraining}\label{sec:results/pretrained}
Preliminary experiments indicated that encoder pretraining is an important step to stabilize the training of the query-based model, especially as the number of query stems grows. The encoder pretraining is done with a common-encoder multi-decoder setup similar to \cite{Watcharasupat2023GeneralizedBandsplitNeural} with a \textsc{vdbo} setup for 100 epochs. The \textsc{vdbo} decoders were discarded and the encoder was used for subsequent experiments. The performance of the pretrained model is shown in \Cref{tab:vdbo},
with performance above oracle ideal ratio mask (IRM) for drums and bass, and on par with HT-Demucs for vocals.\footnote{All coarse-level results for oracle methods, HT-Demucs, and Spleeter were recomputed only on the test set using song-wise results from \href{https://github.com/moises-ai/moises-db/tree/main/benchmark}{github.com/moises-ai/moises-db}. The song-wise results were missing for five of the songs (as of 6 April 2024), two of these belong in the test set, thus the aggregates were computed over 46 songs instead of 48 songs.}

\subsection{Learning to separate from queries} \label{sec:results/vdb}

As a first step to verify the query-based ability of the model, a Banquet model is trained to extract only \textit{lead female singer}, \textit{lead male singer}, \textit{drums}, and \textit{bass} stems, referred to as the \textsc{q:vdb} setup. We experimented with training from scratch, using a frozen pretrained encoder (FE), and using a trainable pretrained encoder (TE). 
While the frozen-encoder setup did not demonstrate any sign of overfitting during the training, the trainable-encoder system demonstrated (very slight) overfitting. As a result, an additional setup with data augmentation (DA) was attempted with the trainable encoder setup, using simple stem-wise within-song random gain (up to $\pm$\SI{6}{dB}), random time shifting, polarity inversion, and channel swapping.

The results are shown in \Cref{tab:vdb}. All three variants with pretrained encoder provided better performance than the model trained from scratch, except for drums in the trainable-encoder model without DA being \SI{0.1}{dB} lower. 
Thus, for all subsequent experiments, the encoder is always pretrained. Without DA, there was no clear benefit to unfreezing the encoder. However, in a trainable encoder system with DA, slight to moderate improvements were observed across all but the male vocal stem. Note, however, that allowing full-model training significantly increases the number of trainable parameters from 13.5 M to 24.9 M thus the computational cost and training time also increases accordingly. The performances of the drums and bass stems are on par or better than the dedicated-stem setup in \Cref{tab:vdbo}. Generally, the models perform better on female vocals than on male vocals. 
\ackedalex{Do we have query-based results for \textsc{vdbo}? It seems to me that these are missing for the narrative. Otherwise you jump directly from multi-decoder \textsc{vdbo} to single-decoder query driven.}

\newcolumntype{C}{>{\centering\arraybackslash}X}
\begin{table}[t]
    \scriptsize
    \caption{Median SNR of the models trained on the \textsc{vdbo} setup, evaluated on the test set of MoisesDB.\ackedalex{aren't there some bold values missing?}} \hskip2pt
    \centering
    
    \begin{tabularx}{\columnwidth}{XS[table-format=-2.1]S[table-format=-2.1]S[table-format=-2.1]S[table-format=-2.1]}
        \toprule
        \textbf{Model} \hfill \textbf{Median SNR (dB):} & {\textbf{Vocals}} & {\textbf{Drums}} & {\textbf{Bass}} & {\textbf{Other}} \\
        \midrule
        Bandit \cite{Watcharasupat2023GeneralizedBandsplitNeural} & \bfseries 9.1 & 9.9 & 10.6 & 6.4 \\
        HT-Demucs \cite{Rouard2023HybridTransformersMusic} & \bfseries 9.1 & \bfseries 11.0 & \bfseries 12.2 & \bfseries 7.3 \\
        Spleeter \cite{Hennequin2020SpleeterFastEfficient} & 7.4 & 6.6 & 6.8 & 5.0 \\
        \midrule
        Oracle IRM & 10.3 & 9.2 & 8.8 & 7.6\\
        \bottomrule
    \end{tabularx}
    \vspace{-2ex}
    \label{tab:vdbo}
\end{table}

\begin{table}[t]
    \centering
    \caption{Median SNR of Banquet models on the \textsc{q:vdb} setup, evaluated with different-song queries.\protect\footnotemark}\hskip2pt
    \scriptsize
    \begin{tabularx}{\columnwidth}{cCC
    S[table-format=-2.1]
    S[table-format=-2.1]
    S[table-format=-2.1]
    S[table-format=-2.1]}
        \toprule
        \textbf{Pretrained Enc.} &\textbf{FE} & \textbf{DA} & {\textbf{Female Vox}} & {\textbf{Male Vox}} & {\textbf{Drums}} & {\textbf{Bass}} \\
        \midrule
        N & N & N & 8.3 & 7.2 & 9.4 & 9.4 \\
        Y & Y & N & 9.8 & 7.6 & 9.9 & 10.2 \\
        Y & N & N & 9.8 & \bfseries 8.0 & 9.3 & 9.8 \\
        Y & N & Y & \bfseries 10.2 & \bfseries 8.0 & \bfseries 10.1 & \bfseries 10.8 \\
        \bottomrule
    \end{tabularx}
    \vspace{-2ex}
    \label{tab:vdb}
\end{table}

\begin{table}[t]
    \centering
    \caption{Coarse-level performance of the Banquet models with different-song queries on the \textsc{q:vdbgp} setup}
    \hskip2pt
    \scriptsize

\begin{tabularx}{\columnwidth}{
    X
    ccc
    S[table-format=2.1]
        S[table-format=2.1]
        S[table-format=2.1]
    S[table-format=2.1]
        S[table-format=2.1]
        S[table-format=2.1]
}
    \toprule
    \textbf{Model} &
    \textbf{FE}  & 
    \textbf{DA} &
    \textbf{BS} & 
    {\textbf{Vox}} &
    {\textbf{Lead Vox}} &
    {\textbf{Drums}} &
    {\textbf{Bass}} &
    {\textbf{Guitar}} &
    {\textbf{Piano}} \\
\midrule
Banquet & Y & N & N & \bfseries 8.0 & 7.9 & 9.8 & 10.5 & 2.3 & 0.8 \\
& &  & Y & 7.9 & 7.7 & 9.6 & 10.5 & 2.2 & 0.9 \\
\cmidrule{2-10}
&N & N & N & 7.4 & \bfseries 8.0 & 9.6 & 10.6 & 3.0 & 2.3 \\
& &  & Y & 7.6 & 7.7 & 9.3 & 10.2 & 2.9 & \bfseries 2.5 \\
\cmidrule{3-10}
& & Y & N & 7.8 & 7.9 & \bfseries 10.1 & 10.9 & 3.2 & 2.2 \\
& &  & Y & 7.6 & 7.9 & 9.5 & \bfseries 11.0 & \bfseries 3.3 & \bfseries 2.5 \\
 \midrule
\multicolumn{4}{l}{HT-Demucs (\textsc{vdbgpo})} & 
    \bfseries 8.9 & \multicolumn{1}{c}{---} & \bfseries 11.6 & \bfseries 12.4 & 2.4 & 1.7 
\\
\multicolumn{4}{l}{Spleeter (\textsc{vdbpo})} & 
    7.0 & \multicolumn{1}{c}{---} & 6.9 & 6.7 & \multicolumn{1}{c}{---} & 0.7
\\
\midrule
\multicolumn{4}{l}{Oracle IRM} & 10.0 & \multicolumn{1}{c}{---} & 9.6 & 7.8 & 5.2 & 5.0\\
\bottomrule
\end{tabularx}
    {\footnotesize Bold: best Banquet model \textbf{and/or} best non-oracle model.}
    \vspace{-3ex}
    \label{tab:vdbgp-merged}
    \ackedalex{Why are there cols with two bold entries?}
\end{table}

\footnotetext[6]{\ackedalex{is this the right footnotenumber?}Median results for the same-song query and different-song query are within \SI{0.2}{\decibel} of each other.}

\subsection{Extending to guitar and piano}\label{sec:results/vdbgp}

Amongst systems that tackled MSS beyond four stems, the next two stems beyond \textsc{vdbo} are usually guitar and piano, due to their high prevalence within pop/rock music. The set of possible queries is thus extended from \textsc{q:vdb} to also include \textit{acoustic guitar}, \textit{clean electric guitar}, \textit{distorted electric guitar}, \textit{grand piano}, and \textit{electric piano}. 
This is referred to as the \textsc{q:vdbgp} setup. Due to the significantly lower number of available training data for guitar and piano stems, we also experimented with a balanced sampling (BS) strategy. In this strategy, a random stem is first chosen as the target stem, then a random song containing that stem is chosen. The remainder of the sampling process is the same as the default. This strategy ensures that every stem has a similar number of training pairs, but distorts the ``natural'' distribution of stem occurrences. 

For comparability with existing systems, the inference outputs of fine-level stems in this setup were added together to form their respective coarse-level predictions.\footnote{The ground truth signals for are the full coarse-level tracks, e.g. \textit{vocals} ground truth include contributions from \textit{background vocals} even if we do not have \textit{background vocals} in the predictions.} 
Coarse-level results are shown in \Cref{tab:vdbgp-merged}. 
Fine-level results for trainable-encoder models are shown in \Cref{tab:vdbgp}.

At the coarse level, most variants of Banquet continue to perform above the oracle IRM for drums and bass. With the default-sampling trainable encoder systems, the Banquet performed better than HT-Demucs on guitar and piano. Without DA, balanced sampling generally did not lead to consistent improvements for guitar and piano. With balanced sampling and DA on a trainable-encoder model, however, slight gains in median SNRs of guitar and piano were observed, albeit at the cost of vocals and drum SNRs.

At the fine level, the model performance follows a similar trend to that of the coarse level. Drums and bass continue to perform above the oracle IRM, while both lead vocals performed close to the IRM. Guitar and piano performances are still well below IRM. Interestingly, it appears that querying with excerpts from the same or different track did not affect the model performance for most stems except for electric piano. This is likely due to both the small sample size of electric piano limiting generalizability, and the highly diverse set of possible timbres thus the intertwined nature of both the query embedding and the target audio with other keyboard instruments. 
The ability of the model to query with stems from different tracks is a double-edged sword, however, since this also means that the model is somewhat insensitive to fine differences in timbre between different renditions of the ``same'' instruments. This could potentially limit its usefulness when applied to a scenario where multiple target stems have very similar timbres.

\begin{table*}[t]
    \centering
    \caption{Model performance on the \textsc{q:vdbgp} setup fine-level stems.}\hskip2pt
    \scriptsize
    \setlength{\tabcolsep}{1pt}
    \begin{tabularx}{\textwidth}{
    CCCC@{\hspace{6pt}}
    S[table-format=2.1]
        S[table-format=2.1]
        S[table-format=2.1]
        @{\hspace{6pt}}
    S[table-format=2.1]
        S[table-format=2.1]
        S[table-format=2.1]
        @{\hspace{6pt}}
    S[table-format=2.1]
        S[table-format=2.1]
        S[table-format=2.1]
        @{\hspace{6pt}}
    S[table-format=2.1]
        S[table-format=2.1]
        S[table-format=2.1]
        @{\hspace{6pt}}
    S[table-format=2.1]
        S[table-format=2.1]
        S[table-format=2.1]
        @{\hspace{6pt}}
    S[table-format=2.1]
        S[table-format=2.1]
        S[table-format=2.1]
        @{\hspace{6pt}}
    S[table-format=2.1]
        S[table-format=2.1]
        S[table-format=2.1]
        @{\hspace{6pt}}
    S[table-format=-2.1]
        S[table-format=2.1]
        S[table-format=2.1]
        @{\hspace{6pt}}
    S[table-format=-2.1]
        S[table-format=2.1]
        S[table-format=2.1]
}
    \toprule
     &
    &&&
    \multicolumn{3}{c}{\textbf{Female Vox}} & 
    \multicolumn{3}{c}{\textbf{Male Vox}} & 
    \multicolumn{3}{c}{\textbf{Drums}} & 
    \multicolumn{3}{c}{\textbf{Bass}} & 
    \multicolumn{3}{c}{\textbf{Acoust. Gtr.}} & 
    \multicolumn{3}{c}{\textbf{Clean E. Gtr.}} & 
    \multicolumn{3}{c}{\textbf{Dist. E. Gtr.}} & 
    \multicolumn{3}{c}{\textbf{Grand Piano}} & 
    \multicolumn{3}{c}{\textbf{E. Piano}}\\
    \cmidrule(lr){5-7}
    \cmidrule(lr){8-10}
    \cmidrule(lr){11-13}
    \cmidrule(lr){14-16}
    \cmidrule(lr){17-19}
    \cmidrule(lr){20-22}
    \cmidrule(lr){23-25}
    \cmidrule(lr){26-28}
    \cmidrule(lr){29-31}
    \textbf{FE}  & 
    \textbf{DA} &
    \textbf{BS} & 
    \textbf{SSQ} &
    {\textbf{Q1}} &
        {\textbf{Q2}} &
        {\textbf{Q3}} &
    {\textbf{Q1}} &
        {\textbf{Q2}} &
        {\textbf{Q3}} &
    {\textbf{Q1}} &
        {\textbf{Q2}} &
        {\textbf{Q3}} &
    {\textbf{Q1}} &
        {\textbf{Q2}} &
        {\textbf{Q3}} &
    {\textbf{Q1}} &
        {\textbf{Q2}} &
        {\textbf{Q3}} &
    {\textbf{Q1}} &
        {\textbf{Q2}} &
        {\textbf{Q3}} &
    {\textbf{Q1}} &
        {\textbf{Q2}} &
        {\textbf{Q3}} &
    {\textbf{Q1}} &
        {\textbf{Q2}} &
        {\textbf{Q3}} &
    {\textbf{Q1}} &
        {\textbf{Q2}} &
        {\textbf{Q3}} \\
\midrule
N & N & N & N & 5.5 & 9.6 & 13.2 & 6.7 & \bfseries 7.9 & 10.0 & 8.0 & 9.6 & 11.6 & 7.9 & 9.9 & 12.0 & 0.9 & \bfseries 1.8 & 3.6 & 0.2 & 0.7 & 2.4 & 0.9 & 2.4 & 5.3 & 0.7 & 2.3 & 2.9 & 0.0 & 0.6 & 0.7 \\
 &  &  & Y & 5.6 & 9.6 & 13.2 & 6.7 & \bfseries 7.9 & 10.0 & 8.0 & 9.6 & 11.6 & 7.9 & 9.9 & 12.0 & 0.9 & \bfseries 1.8 & 3.7 & 0.2 & 0.9 & 2.6 & 0.9 & 2.4 & 5.3 & 0.7 & 2.2 & 3.0 & 0.0 & 0.8 & 1.5 \\
  \cmidrule(){3-31}
 &  & Y & N & 6.1 & 9.6 & 13.1 & 6.8 & 7.7 & 9.7 & 7.8 & 9.3 & 11.3 & 7.6 & 10.0 & 11.5 & 0.8 & \bfseries 1.8 & 3.6 & 0.2 & 0.8 & 2.5 & 1.0 & 2.5 & \bfseries 5.4 & \bfseries 0.8 & 2.5 & 3.1 & -0.1 & 0.7 & 0.8 \\
 &  &  & Y & 6.1 & 9.6 & 13.1 & 6.8 & 7.7 & 9.7 & 7.8 & 9.3 & 11.3 & 7.6 & 10.0 & 11.5 & 0.8 & \bfseries 1.8 & 3.7 & -0.0 & 0.9 & 2.7 & \bfseries 1.2 & 2.5 & \bfseries 5.4 & \bfseries 0.8 & 2.5 & 3.1 & -0.6 & 0.8 & 1.8 \\
 \cmidrule(){2-31}
 & Y & N & N & 5.5 & \bfseries 10.1 & 13.0 & \bfseries 6.9 & \bfseries 7.9 & \bfseries 10.2 & \bfseries 8.5 & \bfseries 10.1 & \bfseries 12.3 & \bfseries 8.4 & \bfseries 10.7 & \bfseries 13.2 & \bfseries 1.2 & 1.7 & 4.5 & 0.2 & 0.9 & \bfseries 3.0 & 0.9 & 2.8 & 4.7 & \bfseries 0.8 & \bfseries 2.8 & \bfseries 3.2 & 0.1 & 0.5 & 0.9 \\
 &  &  & Y & 5.5 & \bfseries 10.1 & 13.1 & \bfseries 6.9 & \bfseries 7.9 & \bfseries 10.2 & \bfseries 8.5 & \bfseries 10.1 & \bfseries 12.3 & \bfseries 8.4 & \bfseries 10.7 & \bfseries 13.2 & \bfseries 1.2 & 1.7 & \bfseries 4.6 & 0.2 & \bfseries 1.1 & 2.7 & 0.9 & 2.8 & 4.7 & \bfseries 0.8 & 2.4 & 3.1 & -0.1 & 0.6 & 0.9 \\
 \cmidrule(){3-31}
 &  & Y & N & 5.5 & \bfseries 10.1 & \bfseries 13.5 & 6.5 & 7.8 & 10.0 & 8.3 & 9.5 & 11.8 & \bfseries 8.4 & 10.3 & 12.1 & 1.1 & 1.7 & 3.9 & 0.0 & 0.4 & 2.7 & 0.9 & \bfseries 3.0 & 4.9 & \bfseries 0.8 & 2.6 & \bfseries 3.2 & 0.2 & 0.5 & 0.9 \\
 &  &  & Y & 5.5 & \bfseries 10.1 & \bfseries 13.5 & 6.5 & 7.8 & 10.0 & 8.3 & 9.5 & 11.8 & 7.8 & 10.3 & 12.1 & 1.0 & 1.7 & 3.9 & \bfseries 0.3 & 0.6 & 2.7 & 0.6 & \bfseries 3.0 & 4.8 & \bfseries 0.8 & 2.5 & \bfseries 3.2 & \bfseries 0.6 & \bfseries 0.9 & \bfseries 2.1 \\


\bottomrule
\end{tabularx}
    {\footnotesize FE: frozen encoder, DA: data augmentation, BS: balanced sampling, SSQ: same-song query, Q1: lower quartile, Q2: median, Q3: upper quartile}
    \vspace{-3ex}
    \ackedalex{Pooh. That's a bit much. Could you reintroduce FE, DA, BS, and SSQ? Also, what is Q1--3? Can this table maybe be color coded?}
    \label{tab:vdbgp}
\end{table*}

\subsection{Extending beyond guitar and piano}\label{sec:results/ev}

The results for the \textsc{q:vdbgp} setup demonstrated that the model is able to learn to extract 5 additional stems. In this experiment, we extend the set of possible queries to include all remaining stems with at least one data point per fold\ackedalex{what is sufficient?}:
\textit{effects}, 
\textit{pitched percussion}, 
\textit{organs \& electronic organs}, 
\textit{synth pad}, \textit{synth lead}, 
\textit{string section}, 
\textit{brass}, and
\textit{reeds}. Additionally, \textit{bass} is now broken up into \textit{bass guitar} and \textit{bass synth}.
This is referred to as the \textsc{q:all} setup. 
Although these are all fine-level stems as defined by MoisesDB,  some of these classes are more specific than others.
For example, \textit{brass} is a fine-level stem despite possibly including trumpets, trombones, horns, and tuba. 
The experimental setups are similar to that of Setup~B.\footnote{BS and DA models for \textsc{q:all} were significantly more unstable during training than for the \textsc{q:vdbgp} setup, despite being identical architecturally. When this happens, we discard the collapsed model and restart the training from scratch until we have a model that completes the entire training run with nonsilent output for most stems. No TE+DA+BS system was stable enough to finish the training run without collapse.}


\begin{figure}[t]
    \centering
    \includegraphics[alt={A boxplot showing performance of different variants of the Banquet models. Read discussion text for specific details.}, width=0.9\columnwidth]{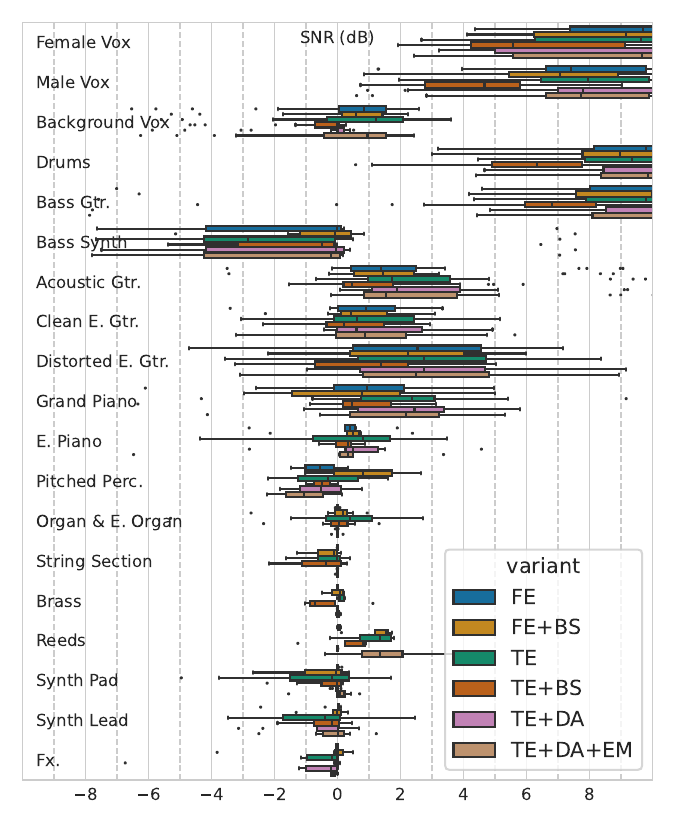}
    \vspace{-3ex}
    \caption{Performance of the Banquet models with same-song queries on \textsc{q:all} fine-level stems}
    \label{fig:setc}
\end{figure}

The same-song query results\footnote{Note that when the FE and the TE+DA systems have SNR concentrated at \SI{0}{\decibel} for the long-tail stems, these are indicators of the model outputting very soft, practically silent output. In general, a model yielding negative SNR for a particular stem might be more desirable than a model that has collapsed for a particular stem.} for the models trained in \textsc{q:all} are shown in \Cref{fig:setc}. 
The performances of the model trained on \textsc{q:all} on the stems from the \textsc{q:vdbgp} setup are similar to those in \Cref{tab:vdbgp},
with the exception of the significant drop in performance for the balanced-sampled trainable-encoder model. Amongst the newly added stems, there are significant variations in performance, but they are all still very weak in terms of SNR, with no sample performing above \SI{5}{dB} SNR. For organs, background vocals, and both synth stems, the trainable-encoder model yielded the better upper quartile and maximum performance, but is also very unreliable. Unfortunately, balanced sampling on a trainable encoder model only worsened the performance. 
DA on a trainable-encoder model with default sampling slightly improved the lower quartile performance, but is also accompanied by lower maximum and upper quartile performance. Frozen-encoder system collapsed for most long-tail stems in default sampling, but balanced sampling interestingly was more stable and performed the best for bass synth, pitched percussion, reeds, and brass. Evidently, the classical tradeoffs are at play here; allowing the model more flexibility with a trainable encoder also comes with a higher risk of model collapse or unreliable performance. More surprisingly, the fact that even a frozen encoder trained on a \textsc{vdbo} setup was able to function at all beyond \textsc{q:vdb} indicates that the embedding space of a Bandit encoder already contains information that is partially generalizable beyond \textsc{vdbo}, as also observed in \cite{Watcharasupat2023GeneralizedBandsplitNeural}.

The results of the long-tail stems are somewhat unsurprising given that the genre distribution in MoisesDB skewed heavily toward pop, rock, and singer-songwriter. In addition to the low track counts, these long-tail instruments also tend to have infrequent active segments and relatively softer levels within a song. In fact, of the long-tail stems, reeds and pitched percussion are the only ones with median RMS above \SI{-35}{dBFS}. Analysis of the SNR distribution shows that the model performance is quite correlated to the track-level RMS of the target signal (Spearman's $\rho$ between \num{0.78} and \num{0.81}). This is likely due to a combination of low data availability and the inherent difficulty associated with cleanly extracting these ``supporting'' stems when there are significant spectral overlaps from more prominent co-occurring stems. In light of the recently published analysis in  \cite{Jeon2024WhyDoesMusic}, we may have been too conservative with our DA setup. In particular, we made a conscious choice to only perform gain augmentation close to the original levels, instead of significantly amplifying softer stems. Whether the latter may improve the result at all will have to be addressed in future work. Moreover, given that \cite{Lin2021UnifiedModelZeroshot} saw partial success with the predominantly classical instrumentation of URMP, there may also be an opportunity for a much more aggressive cross-dataset DA.

%

\ackedalex{Your results paragraphs are mostly descriptive; when you go into more detail it's mostly about the training setup. I wonder whether you can add more insights here in general. Possibly more about instrument characteristics, maybe genre dependencies, dependency on data distribution, anything we can speculate about the query? That's why I was asking for a separate discussion. Also, I think it's important to explain better the relationship with the sota to make sure you present your system in a convincing way even if the results do not look better at first glance.}

\section{Conclusion}
In this work, Banquet, a stem-agnostic single-decoder query-based source separation system was proposed to address MSS beyond the \textsc{vdbo} stems.
At 24.9~M trainable parameters, this highly modularized model with a single stream of information flow provided strong performance for vocals, drums, and bass; outperformed significantly more complex HT-Demucs on guitar and piano; and provided a proof-of-concept for extractions of additional long-tail and/or fine-grained stems at no additional complexity. While there remains room for improvements for long-tail stems with low data availability, this work demonstrated the opportunity for further research on single-decoder systems toward supporting a large and diverse set of stems.

\ifdoubleblind\else
\section{Acknowledgments}

This work was supported by a Cyber-Infrastructure Resource Award from the Institute for Data Engineering and Science (IDEaS), Georgia Institute of Technology. 

K.~N.~Watcharasupat was separately supported by the American Association of University Women (AAUW) International Fellowship, and the IEEE Signal Processing Society Scholarship Program. 

The authors would like to thank Yiwei Ding and Chih-Wei Wu for their assistance with the project.

\section{Ethics Statement}

Machine learning-based systems are inherently data-dependent. Our models rely both directly and indirectly on datasets with inherent imbalance, not only in terms of instrument and genre distribution, but also in terms of cultural origins. As a result, our system inevitably inherit some bias and may not perform on music that have not been well represented in the training data. The authors acknowledge this important limitation and are committed to continue exploring approaches to correct these biases, both in terms of data acquisition and algorithmic development. 



\fi

\small
\bibliography{references}

\end{document}